\documentclass[aps,prd,twocolumn,nofootinbib,superscriptaddress,floatfix]{revtex4-1}

\usepackage{graphicx}
\usepackage{dcolumn}
\usepackage[hypertex]{hyperref}

\newcommand{\nn}{\nonumber}
\newcommand{\beq}{\begin{equation}}
\newcommand{\eeq}{\end{equation}}
\newcommand{\beqa}{\begin{eqnarray}}
\newcommand{\eeqa}{\end{eqnarray}}

\newcommand{\GeV}{\text{GeV}}

\newcommand{\MeV}{\text{MeV}}

\def\mbos{m_b^{1S}}
\def\mcos{m_c^{1S}}
\def\mcmc{ \overline m_c(\overline m_c) }
\def\Gcc{\Gamma_{\!c\bar c}}
\def\mref{m_{\text{ref}}}

\def\lqcd{\Lambda_{\rm QCD}}

\begin{document}

\title{\boldmath Constraining weak annihilation using semileptonic $D$ decays}

\author{Zoltan Ligeti}
\affiliation{Ernest Orlando Lawrence Berkeley National Laboratory,
  University of California, Berkeley, CA 94720}

\author{Michael~Luke}
\affiliation{Department of Physics, University of Toronto,
  60 St.\ George Street, Toronto, Ontario, Canada M5S 1A7}

\author{Aneesh V.\ Manohar}
\affiliation{Department of Physics, University of California at San Diego, 
  9500 Gilman Drive, La Jolla, CA 92093}

\begin{abstract}

The recently measured semileptonic $D_s$ decay rate can be used to constrain
weak annihilation (WA) effects in semileptonic $D$ and $B$ decays.  We revisit
the theoretical predictions for inclusive semileptonic $D_{(s)}$ decays using a
variety of quark mass schemes.  The most reliable results are obtained if the
fits to $B$ decay distributions are used to eliminate the charm quark mass
dependence, without using any specific charm mass scheme.  Our fit to the
available data shows that WA is smaller than  commonly assumed. There is no
indication that the WA octet contribution (which is better constrained than the
singlet contribution) dominates. The results constrain an important source of
uncertainty in the extraction of $|V_{ub}|$ from inclusive semileptonic $B$
decays.

\end{abstract}

\maketitle

An intriguing hint of a possible conflict in the $B$-factory data, which may be
a sign of physics beyond the standard model, is the roughly $2\sigma$ difference
between the value of $|V_{ub}|$ obtained from a global fit to the CKM parameters
using $\sin2\beta$ extracted from the $CP$ asymmetry in $B\to \psi\, K_S$ and
related modes, and $|V_{ub}|$ measured directly from inclusive semileptonic
$B\to X_u \ell\bar\nu$ decays, calculable using an operator product expansion
(OPE)~\cite{Chay:1990da,[][{ [Erratum, {\sl ibid.}\ B {\bf 297} (1993)
477]}.]Bigi:1992su,Bigi:1993fe,Manohar:1993qn}. At order $\lqcd^3/m_b^3$ in the
OPE, four-quark operators, the so-called weak annihilation (WA) terms, give a
significant contribution in a phase space region which affects all inclusive
$|V_{ub}|$ measurements to some extent.  Hence, a reliable estimate of the WA
contribution is necessary to determine whether there is, in fact, a conflict
with standard model predictions.

The WA contribution to the total $B\to X_u \ell\bar\nu$ decay
rate~\cite{Bigi:1993bh,Voloshin:2001xi}, and to the charged lepton (or neutrino)
energy spectrum~\cite{Leibovich:2002ys} are calculable  in terms of the matrix
elements of local four-quark operators. However, there is so far no
first-principles derivation of the WA contribution to the double or triple
differential spectra. For this reason, the extraction of the WA contribution
from the differential $B\to X_u \ell\bar\nu$ spectra is model dependent, and any
model-independent bound on the magnitude of the WA matrix elements is
important.  It was pointed out by Voloshin~\cite{Voloshin:2001xi} that the same
matrix elements that enter $B\to X_u \ell\bar\nu$ decay can be constrained by
the semileptonic rate difference between $D$ and $D_s$ mesons, since the $B$ and
$D$ matrix elements are related by heavy quark symmetry. In this paper we revist
the theoretical calculations of semileptonic $D$ decays and extract bounds on
the WA contribution to $D$ and $B$ decays.

At order $\lqcd^3/m_c^3$, there are dimension-6 four-quark operators in the
OPE for the semileptonic $D$ decay rate, the WA operators,
\beqa\label{dim6ops}
O_{V-A}^{(q)} &=& (\bar c_v \gamma^\mu \, q_L)\, 
  (\bar q_L\, \gamma_\mu \, c_v) , \nn\\
O_{S-P}^{(q)} &=& (\bar c_v \, q_L)\, (\bar q_L\, c_v) ,
\eeqa
where $q=s,d$ and $c_v$ is the heavy quark effective theory charm quark field. 
The matrix elements $B_{1,2}^{(q,i)}$ of these operators are defined by
\beqa\label{bagparameters}
\langle D_i | O_{V-A}^{(q)} | D_i \rangle
  &=& \frac18\, f_D^2\, m_D\, B_1^{(q,i)} , \nn\\
\langle D_i | O_{S-P}^{(q)} | D_i \rangle
  &=& \frac18\, f_D^2\, m_D\, B_2^{(q,i)} \,.
\eeqa
where $i=u,d,s$ labels the flavor of the light quark in the $D$ meson.
Compared to the dimension-3, 5, and other dimension-6 operators, the matrix
elements of these operators are enhanced by $16\pi^2$, and contribute to the
semileptonic decay widths of the three $D$ mesons as~\cite{Voloshin:2001xi}
\begin{equation}\label{volo}
\frac{\Gamma_{\rm WA}^{(D_i)}}{\Gamma_0} = \sum_{q=s,d} 
  \frac{f_D^2\, m_D\, |V_{cq}|^2}{m_c^3}\, 16\pi^2 
  \Big( B_2^{(q,i)} - B_1^{(q,i)} \Big) ,
\end{equation}
relative to the semileptonic width at lowest order in the OPE and at tree-level
in $\alpha_s$,
\beq\label{Gamma0}
\Gamma_0(m_c) \equiv  \frac{G_F^2\,  m_c^5}{192\pi^3}\,.
\eeq
If one assumes factorization and the vacuum saturation approximation, then the
WA contribution vanishes, since $B_1 = B_2 = 1$ or $0$ depending on whether or
not $q$ is the same as $i$. Deviations from the factorization ansatz are usually
estimated at the 10\% level~\cite{Voloshin:2001xi}. 

The above analysis also holds for $B$ decays, with the replacement of $D$ meson
quantities by the corresponding $B$ meson ones. The matrix elements in the $c$
and $b$ sectors are related by heavy quark symmetry. The same matrix element
estimate for $B$ decays (i.e., $|B_1 - B_2| = 0.1$), along with $f_B \approx
200\,$MeV~\cite{[][{ [see also
\url{http://www.latticeaverages.org}]}.]Laiho:2009eu}, implies that the
four-quark operators contribute  $\sim 3$\% to the total $B\to X_u \ell\bar\nu$
rate, making it difficult to accurately determine the WA contribution from $B$
decays. However, the WA contribution to $D$ decay is formally $(m_b/m_c)^3$
enhanced relative to $B$ decay, and is comparable to the leading order decay
rate, $\Gamma_0$, due to the $16\pi^2$ enhancement. Thus studying WA effects in
$D$ decay is a good way to constrain the matrix elements of the four-quark
operators. Note that to determine the WA contribution to $B$ decays at the 1\%
level only requires the four-quark matrix elements to $\sim30$\% accuracy. Even
if this $1/m_c^3$ contribution to $D$ decays is comparable to the leading order
rate, this does not necessarily mean that the $1/m_c$ expansion breaks down,
since the WA contribution is the only $16\pi^2$ enhanced contribution at ${\cal
O}(1/m_c^3)$.

It has long been known that the difference in the semileptonic branching 
ratios~\cite{Amsler:2008zzb,Asner:2009pu}
\beqa\label{slwid12}
{\cal B}(D^0\to X e^+\nu) &=& (6.49 \pm 0.09 \pm 0.11)\,\% \,, \nn\\
{\cal B}(D^+\to X e^+\nu) &=& (16.13 \pm 0.10 \pm 0.29)\,\% \,,
\eeqa
is mainly due to the lifetime difference, and  that the semileptonic widths
$\Gamma(D^0\to X e^+\nu) \approx \Gamma(D^+\to X e^+\nu)$ are equal to within
$3\,\%$. Recently CLEO-c measured~\cite{Asner:2009pu} the $D_s$ semileptonic
branching ratio
\beq\label{slwid3}
{\cal B}(D_s^+\to X e^+\nu) = (6.52 \pm 0.39 \pm 0.15)\,\% \,.
\eeq

The expressions for semileptonic $D_{(s)}\to X\ell\bar\nu$ decays are well known
in the literature.  Schematically,
\beq\label{slrate}
\Gamma_{\rm SL} = \Gcc + \Gamma_{\rm WA} \nn\\
\eeq
where $\Gamma_{\rm WA}$ is defined in Eq.~(\ref{volo}), and
\beqa\label{pert}
\Gcc &=& \Gamma_0 \bigg[ |V_{cs}|^2 \big(
  1 - 8r + 8r^3 - r^4 - 12r^2 \ln r \big) + |V_{cd}|^2 \nn\\
&+& \bigg(\lambda_1+ \frac{{\cal T}_1+3{\cal T}_2}{m_c}\bigg) \frac{1}{2m_c^2}
  - \bigg(\lambda_2+ \frac{{\cal T}_3+3{\cal T}_4}{3m_c}\bigg)\frac{9}{2m_c^2}
  \nn\\
&+& \frac{77 \rho_1 + 27\rho_2}{6m_c^3}
  + {\cal O}\bigg(\alpha_s, \frac{\lqcd^4}{m_c^4}\bigg) \bigg] ,
\eeqa
where $\lambda_{1,2}$ and $\rho_{1,2}$ are the matrix elements of dimension-5 
and 6 two-quark operators, ${\cal T}_{1,2,3,4}$ are the matrix elements of
time-ordered products, and $r = m_s^2/m_c^2$.  These may all be determined  from
fits to various $B$ decay spectra~\cite{Bauer:2004ve, Bauer:2002sh,
Buchmuller:2005zv}.  The complete expression including $\alpha_s$ corrections is
complicated.  In our analysis, we include the radiative corrections to order
$\alpha_s^2$~\cite{vanRitbergen:1999gs} in the $1/m_c^0$ terms, and power
corrections to order $\lqcd^3/m_c^3$~\cite{Gremm:1996df,Blok:1994cd}.  For the
leading term in the OPE, we include the effect of a nonzero strange quark mass,
since it affects the semileptonic width by $\sim 6$\%, and set $m_s \to 0$
elsewhere.  For nonzero $r$, the $\rho_1$ contribution to $\Gcc$ has a $\log r$
divergence as $r \to 0$.  In the OPE with $r = 0$, this divergence is
effectively absorbed into the matrix elements of WA operators.  Including the WA
contribution converts the $q^2$ spectrum into a plus
distribution~\cite{Ligeti:2007sn}, which integrates to zero and gives the
$77\rho_1/(6 m_c^3)$ contribution to the total semileptonic rate in
Eq.~(\ref{pert}).

The terms in $\Gcc$ are all independent of the flavor of the spectator quark in
the heavy meson, and so give equal contributions to $\Gamma_{\text{SL}}$ for all
three $D$ (or $B$) mesons. The leading term in $\Gamma_{\rm SL}$ which depends
on the flavor of the spectator quark, and thus produces a difference in the
semileptonic partial widths, is $\Gamma_{\rm WA}$.  $\Gamma_{\rm WA}$ depends
on two independent matrix elements in the flavor $SU(3)$ limit, since the
operators in Eq.~(\ref{dim6ops}) have an $SU(3)$ singlet and octet part, each of
which yield one invariant with the two $D$ fields. We can write the decay rates
of the three $D$ mesons in terms of these two parameters as
\beqa\label{slrateAB}
\!\!\!\!\!\!\Gamma_{\rm SL}^{(D^0)} &=& \Gcc + G_0
  \Big(|V_{cs}|^2\, \frac{a_0-a_8}3 + |V_{cd}|^2\,\frac{a_0-a_8}3 \Big), \\
\!\!\!\!\!\!\Gamma_{\rm SL}^{(D^\pm)} &=& \Gcc + G_0
  \Big(|V_{cs}|^2\, \frac{a_0-a_8}3 + |V_{cd}|^2\,\frac{a_0+2a_8}3 \Big),\nn\\
\!\!\!\!\!\!\Gamma_{\rm SL}^{(D_s)} &=& \Gcc + G_0
  \Big(|V_{cs}|^2\, \frac{a_0+2a_8}3 + |V_{cd}|^2\,\frac{a_0-a_8}3 \Big).\nn
\eeqa
Here we normalized the weak annihilation contribution to the observed
semileptonic $D^{\pm,0}$ decay rate,
\beq\label{G0}
G_0 \equiv  0.16\, {\rm ps}^{-1} \equiv \Gamma_0(\mref) \,,
\eeq
where $\mref = 1.357\, \GeV$ and $a_0$ and $a_8$ are dimensionless
numbers proportional to the singlet and octet matrix elements,
\begin{equation}\label{a08def}
\frac{G_F^2\,m_c^2\,f_D^2\, m_D }{12\pi\, G_0}\, 
  \Big( B_2^{(q,i)} - B_1^{(q,i)} \Big) 
  \equiv \delta_{q,i}\, a_8 + \frac13 (a_0-a_8) .
\end{equation}
The size of $a_{0,8}$ is then (approximately) the fraction of the meson
semileptonic width due to WA. 

The difference $\Gamma_{\rm SL}^{(D^0)}-\Gamma_{\rm SL}^{(D^\pm)}$ is suppressed
by $|V_{cd}/V_{cs}|^2$.  Neglecting this correction, it is straightforward to
extract $a_8$ from the measured difference of the semileptonic widths of the
$D^{0,\pm}$ and the $D_s$, as proposed in
Refs.~\cite{Voloshin:2001xi,Voloshin:2002je}.  This difference combined with any
of the individual semileptonic widths also allows $a_0$ to be extracted, but
this requires a reliable computation of $\Gcc$~\cite{Becirevic:2008us}.  Since
the charm mass is not particularly large compared with nonperturbative QCD
scales, this computation  suffers from both large perturbative and $1/m_c^n$
corrections, limiting the precision with which WA can be studied in charm
decays.  

The leading perturbative corrections to the semileptonic decay widths are given
by a perturbation series multiplying the free-quark decay width $\Gamma_0$ given
in Eq.~(\ref{Gamma0}).  This perturbation series depends on the choice of scheme
for $m_c$, and $m_c$ could be determined from the $B$ decay data using the
method of Ref.~\cite{Hoang:2005zw}.  However, it is well-known that the
perturbation series for this leading term in the OPE is badly behaved when the
rate is expressed in terms of the charm quark pole or $\overline{\rm MS}$
masses.  Including the known results up to order
$\alpha_s^2$~\cite{vanRitbergen:1999gs} and using $\alpha_s =0.35$  gives the
series
\beq\label{poleseries}
\frac{\Gcc}{\Gamma_0\big[m_c^{\rm pole}\big]} = 
1 - 0.269\, \epsilon - 0.360\, \epsilon^2_{\rm BLM} 
  + 0.069\, \epsilon^2 + \dots ,
\eeq
and
\beq\label{MSbarseries}
\frac{\Gcc}{\Gamma_0\big[\mcmc\big]} =
1 + 0.474\, \epsilon + 0.513\, \epsilon^2_{\rm BLM} 
  - 0.142\, \epsilon^2 + \dots ,
\eeq
respectively.  Here $\epsilon \equiv 1$ counts the order in the perturbation
series, and the BLM subscript refers to the $\epsilon^n \beta_0^{n-1}$ terms in
the perturbation series.  [In Eqs.~(\ref{poleseries})~--~(\ref{kinseries}) we
set $m_s\to 0$ for simplicity; this has no effect on our discussion.]  These
series are poorly behaved, and do not appear to converge.

The bad behavior of these perturbation series is understood theoretically from
$b$ decays, and arises from a poor choice for the heavy quark mass. A better
behaved series is obtained by using a threshold mass scheme, such as the $1S$
\cite{Hoang:1998ng,Hoang:1998hm,Hoang:1999ye}, kinetic \cite{Czarnecki:1997sz}
or PS \cite{Beneke:1998rk} mass schemes.  As observed already
in~\cite{Hoang:1998ng}, the perturbation series relating $\Gcc$ to  the $1S$
mass is reasonably well-behaved, and extracting $m_c$ using the method of
Ref.~\cite{Hoang:2005zw}
\beq\label{1Sseries}
\frac{\Gcc}{\Gamma_0\big[\mcos\big]} =
1 - 0.133\, \epsilon - 0.006\, \epsilon^2_{\rm BLM} - 0.017\, \epsilon^2.
\eeq
The series is less well-behaved in the PS or kinetic schemes (defining both
with a 1\,GeV factorization scale).  For the PS scheme we find
\beq\label{PSseries}
\frac{\Gcc}{\Gamma_0\big[m_c^{\rm PS}(1 {\rm GeV})\big]} =
1 + 0.262\, \epsilon + 0.217\, \epsilon^2_{\rm BLM} - 0.079\, \epsilon^2,
\eeq
while for the kinetic scheme, as previously noted~\cite{Kamenik:2009ze}, the
series is considerably worse
\beq\label{kinseries}
\frac{\Gcc}{\Gamma_0\big[m_c^{\rm kin}(1 {\rm GeV})\big]} =
1 + 0.628\, \epsilon + 0.631\, \epsilon^2_{\rm BLM} - 0.126\, \epsilon^2.
\eeq

In addition to the uncertainties in the above series, there will be additional
uncertainties in extracting a charm quark threshold mass from other physical
quantities, such as moments of $B$ decay spectra
\cite{Bauer:2004ve,Bauer:2002sh}.  Since the charm quark mass is an intermediate
quantity which is not required for our analysis, we can minimize this source of
theoretical uncertainty by bypassing any choice of charm mass scheme, and
instead directly relate the semileptonic $D$ decay widths to the values of
$\mbos$ and $\Delta=m_b-m_c$ extracted from a global fit to $B$ decay spectra. 
From Eq.~(\ref{poleseries}) and the relation between the $b$ quark pole and $1S$
masses \cite{Hoang:1998hm}
\beq
{m_b\over \mbos} = 1 + 0.011\, \epsilon + 0.019\, \epsilon^2_{\rm BLM}
  - 0.003\, \epsilon^2 + \dots ,
\eeq
we find the reasonably well-behaved perturbation series
\beq\label{directpert}
\frac{\Gcc}{\Gamma_0\big[\mbos-\Delta\big]}=
1 - 0.075\epsilon - 0.013\, \epsilon^2_{\rm BLM}
  - 0.021\, \epsilon^2 ,
\eeq
using $\mbos=4.7$~GeV, $\Delta=3.4$~GeV, $\alpha_s(m_b)=0.22$, and, as in the
previous expressions, we have continued to set $m_s$ to zero.  We will therefore
use this method to determine the $D$ semileptonic widths theoretically.

The masses $\mbos$ and $m_b-m_c$ and HQET parameters $\lambda_{1,2}$,
$\rho_{1,2}$  and ${\cal T}_{1,2,3,4}$, as well as their correlated
uncertainties, are obtained using a fit to the $B$ decay
spectra~\cite{Bauer:2004ve,Bauer:2002sh}.  The values for $D$ decay are related
to those for $B$ decay by renormalization group evolution between $m_b$ and
$m_c$. $\lambda_1$ is not renormalized due to reparametrization
invariance~\cite{Luke:1992cs}, while $\lambda_2(m_c)= \kappa_c\,
\lambda_2(m_b)$, with $\kappa_c \approx 1.2$. Radiative corrections to the
$1/m_c^3$ terms are computed in
Refs.~\cite{Manohar:1997qy,Bauer:1997gs,Manohar:2010sf}. Since they are small
and were not included in the $B$ decay fits, we neglect them here.

The errors from the fits include the experimental uncertainties, as well as
additional theoretical uncertainties due to neglected higher order corrections,
as given in Ref.~\cite{Bauer:2004ve,Bauer:2002sh}.  We treat the $B$ and $D$
decay calculations as independent. Thus the $B$ decay fit results will be held
fixed (at order $\epsilon^2_{\rm BLM}$) while we vary the order of the $D$ decay
results between tree-level and $\epsilon^2$. While this may be formally
inconsistent, numerically, the $\alpha_s$ and $1/m_Q$ corrections are
significantly larger for $D$ than for $B$ decay.

\begin{figure*}
\includegraphics[bb=18 163 349 476,width=7.1cm]{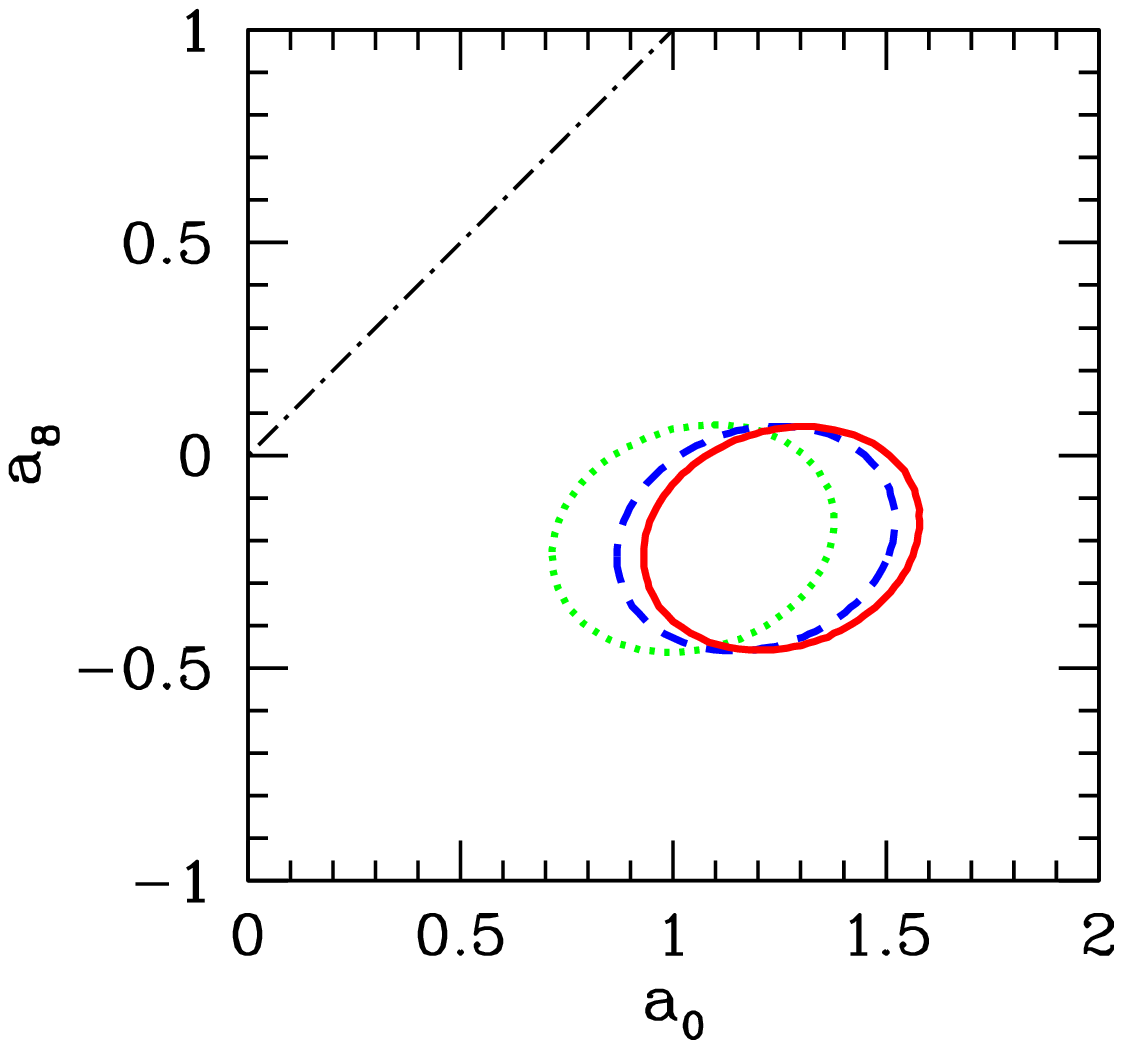}
\hfil\hfil
\includegraphics[bb=18 163 349 476,width=7.1cm]{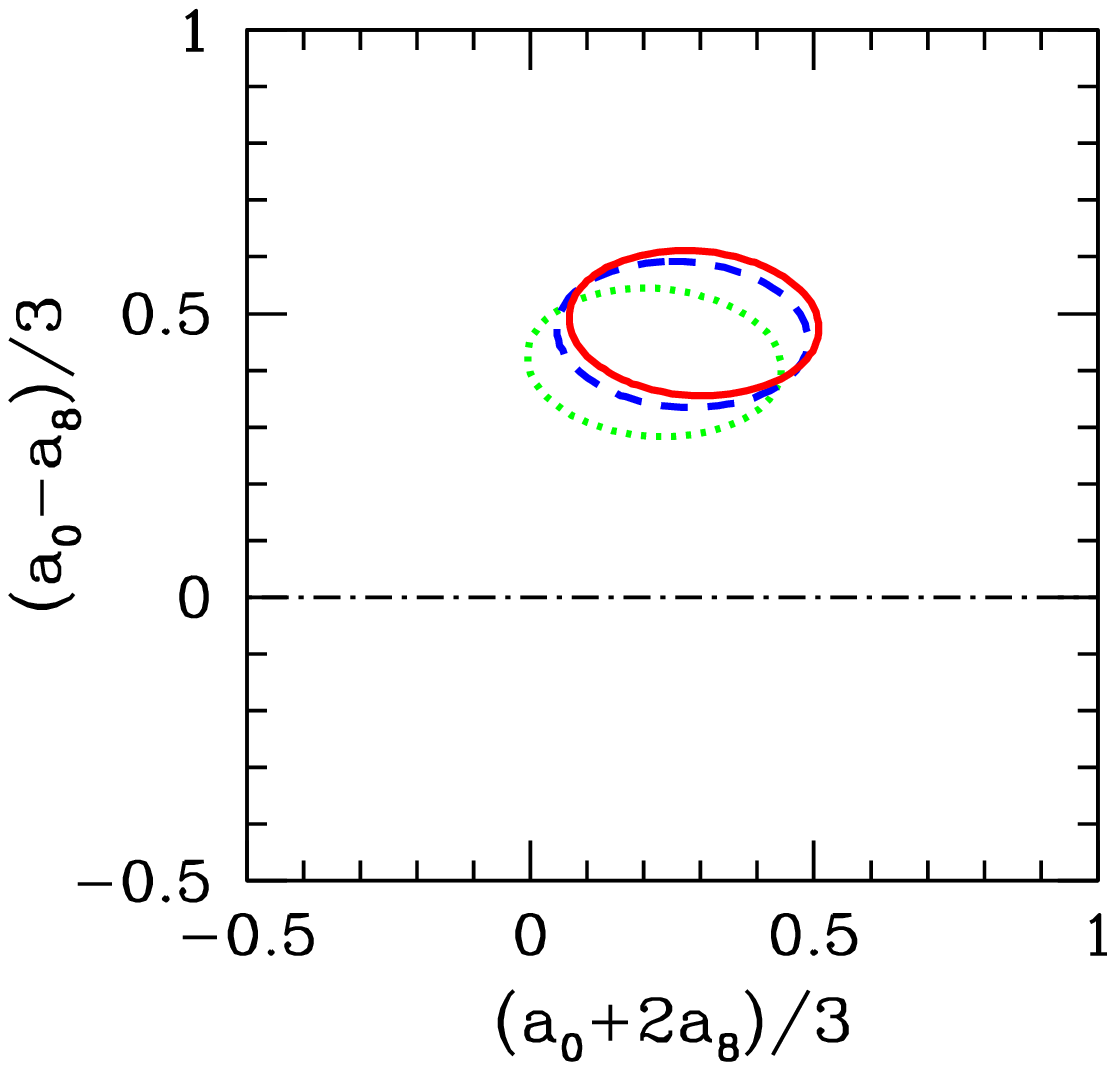}
\caption{The 90\% CL contours for fits at order $\epsilon^0$ (dotted green),
$\epsilon^1$ (dashed blue), and $\epsilon^2$ (solid red). Note that $a_8$ is not
affected by the order in $\epsilon$. The thin dot-dashed black line is the
large-$N_c$ relation, $a_0=a_8$.}
\label{fig:08}
\end{figure*}

Using the value of $\Gcc$ obtained as discussed above, and fitting to the
experimentally measured rates in Eqs.~(\ref{slwid12}) and (\ref{slwid3}) gives
the WA annihilation parameters $a_0$ and $a_8$.  Figure~\ref{fig:08} shows the
90\% CL contours at tree level, order $\epsilon$, and order $\epsilon^2$. The
best fit parameters at order $\epsilon^2$ are
\beqa
a_0 &=& 1.25 \pm 0.15 \,, \qquad \nn\\
a_8 &=& -0.20 \pm 0.12 \,,
\label{a0a8}
\eeqa
where the error is from the order $\epsilon^2$ fit. The series of $\alpha_s^n$
corrections to $\Gcc$ are flavor independent, and lead to a shift in $a_0$
depending on the order in $\epsilon$, but do not affect $a_8$, which can be
determined from $\Gamma_{\rm SL}^{(D_s)} - \Gamma_{\rm SL}^{(D^0)}$.  $\Gcc$
cancels in this difference, so $a_8$ is not affected by the convergence of the
$\alpha_s$ expansion. The shift in $a_0$ between $\epsilon$ and $\epsilon^2$ is
0.06, which is smaller than other uncertainties. The expansion in $1/m_c$  is
also not as rapidly convergent as in the $B$ meson system, so there are
significant uncertainties  which mainly affect $a_0$.  We find that the
$1/m_c^2$ and $1/m_c^3$ terms in Eq.~(\ref{slrate}) contribute roughly $-50\,\%$
and $+35\,\%$ to the semileptonic widths.  These corrections are much  larger
than the corresponding ones for the hadron masses, because of the larger
coefficients of $\lambda_2$, $\rho_1$, and $\rho_2$.  One could estimate the
uncertainty corresponding to these large corrections by including an additional
error of $0.2$ in $a_0$, which is  half the $1/m_c^3$ term. In the $N_c \to
\infty$ limit, the meson sector of QCD has a $U(3)_q \otimes U(3)_{\bar q}$
symmetry~\cite{Veneziano:1979ec} and this implies that $a_0=a_8$ (see, e.g.,
Ref.~\cite{Manohar:1998xv}), which is shown as the black line in
Fig.~\ref{fig:08}.

Neglecting Cabibbo-suppressed terms, the correspondence between our notation and
that of Ref.~\cite{Voloshin:2001xi} is
\beq
a_8 = \frac{m_c^2\, m_D f_D^2}{\mref^5}\, 16\pi^2
  \big( B_2^{ns} - B_1^{ns} \big) ,
\eeq
and the same equation with $a_8 \to a_0$ and the non-singlet $B_{1,2}$ replaced
by the singlet ones.  Taking $f_D \approx 200\,\MeV$~\cite{Laiho:2009eu} gives
$(B_1^{ns} - B_2^{ns}) \approx -a_8 / 4.3 \approx 0.05 \pm 0.03$, which is
somewhat smaller than (although consistent with) the simple estimate $(B_1^{ns}
- B_2^{ns}) \sim 0.1$ in \cite{Voloshin:2001xi,Voloshin:2002je}.

The linear combinations $(a_0+2a_8)/3$ and $(a_0-a_8)/3$ that contribute to the
decay rates in Eq.~(\ref{slrateAB}) are
\begin{eqnarray}
(a_0+2a_8)/3 &=& 0.29 \pm 0.10\,, \nn\\*
(a_0-a_8)/3 &=& 0.48 \pm 0.06\,,
\label{apam}
\end{eqnarray}
where only the fit uncertainty is quoted, as discussed above.  The 90\%
confidence level contours in these variables are shown in Fig.~\ref{fig:08}.
While there are significant uncertainties in the fit result for the WA
contribution in Eqs.~(\ref{a0a8}) and (\ref{apam}), it still has important
implications for $B$ and $D$ decays and the determination of $|V_{ub}|$.

It has often been assumed that the WA term where the light quark in the operator
matches that in the heavy meson is much larger than when the light quarks
differ, i.e., $|a_0+2a_8| \gg |a_0-a_8|$.  Indeed, the central values of our
results suggest that the WA contribution to $B^0$ decay is larger than that to
$B^\pm$ decay.  The WA matrix element in which the light quark field of the
operator is contracted with the spectator quark in the heavy meson is helicity
suppressed by $m_\ell^2/m_c^2$, where $m_\ell$ is the lepton mass, and gives a
contribution of relative order $\lqcd^3 m_\ell^2/m_c^5$ to the decay width. 
Other diagrams, in which the spectator quark is not annihilated by the
four-quark operator, are of relative order $\lqcd^3 /m_c^3$.  In a quark model,
they would contain additional suppression factors from gluon exchange to connect
the spectator light quark with the rest of the diagram, but nothing as small as
$m_\ell^2/m_c^2$.

The $D$ meson lifetimes also depend on the WA matrix elements through both the
semileptonic and non-leptonic decay rates. The non-leptonic rates depend on two
additional color octet operators, and the behavior of the $\alpha_s$
perturbation series is even worse than for the semileptonic case. Neglecting the
color octet matrix elements and $SU(3)$ violation (as before), one would
predict~\cite{Voloshin:2001xi}
\beq
\frac{\Gamma_{\rm SL}^{(D^0)} - \Gamma_{\rm SL}^{(D_s)}}
  {\Gamma_{\rm total}^{(D^0)} - \Gamma_{\rm total}^{(D_s)}}
  = \frac{3}{8\, C_+\, C_-\, \cos^2\theta_C} \approx 0.3 \,,
\eeq
where $C_- = C_+^{-2} = [\alpha_s(m_c) / \alpha_s(m_W)]^{12/25}$, and we have
used $C_- = 1.6$ and $C_+ = 0.8$ for the numerical values. The $D$ branching
ratios  Eqs.~(\ref{slwid12}) and (\ref{slwid3}) and the lifetimes yield $0.07
\pm 0.02$. This shows that there must be some other large contribution to the
nonleptonic decay rates, e.g., large color octet matrix elements, $\alpha_s$
corrections, or higher order $1/m_c$ terms, so the total widths do not provide a
useful bound on $a_{0,8}$.

It is often stated that the difference between the $B^\pm$ and $B^0$
semileptonic rates can be used to constrain the impact of WA on the extraction
of $|V_{ub}|$ from $B \to X_u \ell\bar \nu$ decays.  However, $\Gamma_{\rm
SL}^{(B^\pm)} - \Gamma_{\rm SL}^{(B^0)} \propto a_8$, while individually
$\Gamma_{\rm SL}^{(B^\pm)}$ and $\Gamma_{\rm SL}^{(B^0)}$, which determine
$|V_{ub}|$, depend on both $a_0$ and $a_8$.  We find no evidence that $|a_0| \ll
|a_8|$, so the $\Gamma_{\rm SL}^{(B^\pm)} - \Gamma_{\rm SL}^{(B^0)}$ width
difference will not strongly constrain the WA contribution to $|V_{ub}|$.  While
the uncertainties in our analysis are substantial, it gives strong indication
that the WA contribution to the $B\to X_u\ell\bar\nu$ rate is less than the
$\sim3\%$ estimate~\cite{Voloshin:2001xi} often used. Our conclusions are
unchanged if $SU(3)$ breaking or higher order $1/m_Q$ corrections are included,
since these will only shift the estimate of the WA contribution by $\sim 20$\%
of its value. The $2 \sigma$ discrepancy in $V_{ub}$ mentioned in the
introduction cannot be explained away using WA.

Our results imply that the WA contribution to $B$ decays, which is a factor
$(m_c/m_b)^3 \sim 0.03$ smaller than the corresponding contribution to $D$
decays, is around 1\%.  If we use heavy quark symmetry for the bag parameters
instead of the matrix elements, scaling with $(m_c^2/m_b^2)(f_B^2/f_D^2)$ gives
$2.5\,\%$, still smaller than past estimates.  A recent CDF
measurement~\cite{cdfnote} of the $B$ meson and $\Lambda_b$ baryon lifetimes
also indicates that spectator effects in the $b$ hadron decays may be smaller
than previously thought.

\begin{acknowledgments}

We thank Frank Tackmann for helpful conversations, and C.\ S.\ Park and S.\
Stone for correspondence about Ref.~\cite{Asner:2009pu}.
The work of ZL was supported in part by the Director, Office of Science, Office
of High Energy Physics of the U.S.\ Department of Energy under contract
DE-AC02-05CH11231.  ML was supported in part by the Natural Sciences and
Engineering Research Council of Canada.

\end{acknowledgments}

\bibliography{Dpaper_prd}

\end{document}